\documentclass[useAMS,usenatbib]{mn2e} 
\usepackage{graphicx} 
 
\newcommand{\mnras}{MNRAS}
\newcommand{\aj}{ApJ}
\newcommand{\apj}{ApJ}
\newcommand{\apjs}{ApJ}
\newcommand{\apjl}{ApJ}
\newcommand{\aap}{A\&A}
\newcommand{\aaps}{A\&A}

\newcommand{\gppr}{\stackrel{>}{\scriptstyle \sim}}
\newcommand{\gappr}{\raisebox{-0.4ex}{$\gppr$}}
\newcommand{\lppr}{\stackrel{<}{\scriptstyle \sim}}
\newcommand{\lappr}{\raisebox{-0.4ex}{$\lppr$}}
\newcommand{\Mwd}{\mbox{$M_\mathrm{wd}$}}

\newcommand{\Msun}{\mbox{$\mathrm{M}_{\odot}$}}

\newcommand{\Teff}{\mbox{$T_{\mathrm{eff}}$}}

\newcommand{\kms}{\mbox{$\mathrm{km\,s^{-1}}$}}

\title[Low-mass white dwarfs]{Post-common envelope
  binaries from SDSS-X: The origin of low-mass white dwarfs}
\author[Rebassa-Mansergas et al.]{A.  Rebassa-Mansergas$^1$, A. Nebot
  G\'omez-Mor\'an$^{2,3}$, M.R.  Schreiber$^1$, J. Girven$^4$ \newauthor and
  B.T.G\"ansicke$^4$\\
$^{1}$ Departamento de F\'\i sica y Astronom\'\i a, Universidad de Valpara\'\i so, 
Avenida Gran Bretana 1111, Valpara\'\i so, Chile \\
$^{2}$ Astrophysikalisches Inst. Potsdam, An der Sternwarte 16, 14482,
Potsdam, Germany\\
$^{3}$ CNRS, Observatoire Astronomique, 11 rue de
l'Universite, F-67000 Strasbourg, France\\
$^{4}$ Department of Physics, University of Warwick, Coventry CV4 7AL, UK \\
}

\begin{document}
\date{Accepted 2010. Received 2010; in original form 2010}
\pagerange{\pageref{firstpage}--\pageref{lastpage}} \pubyear{2010}
\maketitle

\begin{abstract}
  We present the  first white dwarf mass distributions  of a large and
  homogeneous sample of post-common envelope binaries (PCEBs) and wide
  white  dwarf-main sequence  binaries (WDMS)  directly  obtained from
  observations.   Both  distributions  are statistically  independent,
  with  PCEBs showing  a clear  concentration of  systems  towards the
  low-mass  end  of  the   distribution,  and  the  white  dwarf  mass
  distribution of wide WDMS binaries  being similar to those of single
  white  dwarfs.  Our results  provide evidence  that the  majority of
  low-mass ($\Mwd  \la 0.5\,\Msun$) white  dwarfs are formed  in close
  binaries.
\end{abstract}


\label{firstpage}

\section{Introduction}

The mass distribution of  hydrogen atmosphere white dwarfs is strongly
clustered    around    an    average   value    of    $\sim0.6\,\Msun$
\citep{koesteretal79-1,         marshetal97-1,         kepleretal07-1,
  holbergetal08-1}, as  predicted by models of  single star evolution.
In addition to the pronounced  peak at $\sim0.6\,\Msun$, a second peak
at   lower  masses,  $\sim0.4\,\Msun$,   has  been   frequently  found
\citep[e.g.][]{bergeronetal92-1,  bragagliaetal95-1,  liebertetal05-1,
  kepleretal07-1}.   Given  the evolutionary  time  scale of  low-mass
single main  sequence stars  that are supposed  to form  such low-mass
($\Mwd \la 0.5\,\Msun$) white  dwarfs significantly exceeds the Hubble
time, the existence  of low-mass white dwarfs has  been interpreted as
the result of strong mass  transfer interactions in binaries.  In this
scenario the  more massive  (primary) star in  a main  sequence binary
fills its Roche-lobe on the  giant branch, and dynamical unstable mass
transfer onto the less massive (secondary) star leads to the formation
of a common  envelope engulfing both the core of  the primary star and
the lower  mass secondary  star.  Orbital energy  released due  to the
shrinkage  of  the binary  orbit  is  supposed  to rapidly  expel  the
envelope, and may  terminate the growth of the He  core of the primary
star before it reaches a sufficient mass for He-ignition.  The outcome
of this close binary evolution is a post-common envelope binary (PCEB)
consistent  of  a (possibly  low-mass  He-core)  white  dwarf and  the
(basically   unaltered)   low-mass   secondary   \citep{paczynski76-1,
  webbink84-1, iben+tutukov86-1}.

The   hypothetical  binary   origin  of   low-mass  white   dwarfs  is
observationally  supported by  the  large fraction  of low-mass  white
dwarfs    in    short    orbital    period   double    white    dwarfs
\citep[e.g.][]{marshetal95-1},  in neutron  star-white  dwarf binaries
\citep[e.g.][]{sigurdssonetal03-1} and in  subdwarf B star-white dwarf
pairs \citep[e.g.][]{maxtedetal02-1}, as well as the existence of some
PCEBs       with       low-mass       white      dwarf       primaries
\citep[e.g.][]{schreiber+gaensicke03-1, nebotetal09-1, pyrzasetal09-1,
  zorotovicetal10-1}.

However, also a fair number of apparently single low-mass white dwarfs
are  known  that  exhibit  neither  radial  velocity  variations,  nor
infrared flux excess, the typical hallmarks of white dwarfs with close
companions  \citep{maxtedetal00-4, napiwotzkietal07-1, kilicetal10-1}.
Possible  explanations  for the  existence  of  these systems  include
supernova type Ia explosions in semi-detached close binaries that blow
away the  envelope of  the companion thus  exposing its  low-mass core
\citep{justhametal09-1},  severe mass-loss on  the first  giant branch
\citep{kilicetal07-1} that may even  lead to the formation of low-mass
white dwarfs with CO-cores \citep{prada-moroni+straniero09-1}, stellar
envelope  ejection  due  to  the  spiral-in  of  close  giant  planets
\citep{nelemans+tauris98-1}, or the merging of two very low-mass white
dwarfs \citep{hanetal02-1}.

So far, conclusive studies testing the close binary origin of low-mass
white dwarfs have  been prevented by the lack  of a sufficiently large
and homogeneous sample  of PCEBs. This is now  rapidly changing thanks
to   the  Sloan   Digital  Sky   Survey  \citep[SDSS,][]{yorketal00-1,
  abazajianetal09-1,   yannyetal09-1}  from  which   $\sim2000$  white
dwarf/main  sequence  (WDMS)   binaries  have  been  spectroscopically
identified         \citep{silvestrietal07-1,        schreiberetal07-1,
  helleretal09-1, rebassa-mansergasetal10-1}.  This population of WDMS
binaries  consists of wide  systems whose  stellar components  did not
interact and thus  evolved like single stars, and  close binaries that
suffered from  dynamically unstable mass transfer,  i.e.  PCEBs. Based
on extensive radial velocity follow-up  of the SDSS WDMS binary sample
\citep{rebassa-mansergasetal07-1,            rebassa-mansergasetal08-1,
  schreiberetal08-1, schreiberetal10-1}, we  demonstrate here that the
mass  distribution of  the white  dwarfs in  PCEBs does  indeed differ
significantly from  that of  white dwarfs that  do not  undergo binary
interactions, containing a large fraction of low-mass white dwarfs.

\section{The sample}
\label{s-data}

SDSS  spectroscopy  has  been   very  efficient  in  identifying  WDMS
binaries:  1602  systems  were   found  in  Data  Release  (DR)\,6  by
\citet{rebassa-mansergasetal10-1}.    Several    hundred   more   were
discovered from  a dedicated WDMS SEGUE \citep[the  SDSS Extension for
  Galactic Understanding and Exploration, see][]{yannyetal09-1} survey
(Nebot  G\'omez-Mor\'an et  al.  2011a,  A\&A  in prep.),  and we  are
currently  compiling   the  final  addition  of   systems  from  DR\,7
(Rebassa-Mansergas et al.  2011 in  prep.), taking the total number of
WDMS binaries in  SDSS to over 2000. The system  parameters of all the
SDSS/SEGUE  WDMS  binaries were  determined  by  decomposing the  SDSS
spectra into  their white dwarf and companion  star contributions, and
subsequently  fitting  the  white  dwarf spectra  with  models.   This
procedure is described in detail in \citet{rebassa-mansergasetal07-1}.
However, intensive tests of our white dwarf fitting routine using SDSS
spectra of single white dwarfs (see Sect.\,\ref{s-comp}) revealed that
we over-estimated the errors on the white dwarf parameters by a factor
$\sim2$. We  have hence  re-fitted all WDMS  binary spectra  to obtain
more realistic  uncertainties of  the white dwarf  parameters. Updated
uncertainties for the WDMS binaries  not studied here will be provided
in Rebassa-Mansergas et al.  (2011) in prep.

\begin{table}
\centering
\caption{\label{t-mass}  White  dwarf masses  for  the  211 SDSS  WDMS
  binaries (76 PCEBs and 135 wide WDMS binary candidates) used in this
  work. In  the last column we  indicate whether the object  is a PCEB
  (1) or a  wide WDMS binary (0).  The complete table  can be found in
  the electronic edition of the paper.}  \setlength{\tabcolsep}{1.1ex}
\begin{small}
\begin{tabular}{cccc}
\hline
\hline
Object &  M$_\mathrm{wd}$[\Msun] & error[\Msun]  & pceb? \\
\hline
  SDSS\,J000453.93+265420.4  & 0.606 & 0.095 &    0 \\
  SDSS\,J000559.87--054416.0 & 0.615 & 0.055 &    0 \\
  SDSS\,J000651.91+284647.1  & 0.590 & 0.048 &    0 \\
  SDSS\,J001749.24--000955.3 & 0.496 & 0.034 &    1 \\
  SDSS\,J003221.86+073934.4  & 0.369 & 0.015 &    1 \\
  SDSS\,J003602.59+070047.2  & 0.584 & 0.019 &    0 \\
        ...          & ...   & ...   & ...   \\
\hline
\end{tabular}
\end{small}
\end{table}

The  vast  majority  of the  SDSS  WDMS  binaries  in our  sample  are
spatially unresolved.  Given the  typical distances of $>100$\,pc, and
assuming a typical resolution of $\simeq1"$ for the SDSS images, these
systems  may hence  have  binary separations  of  up to  many tens  of
astronomical  units   and  orbital  periods  of  up   to  hundreds  of
years. This  implies that  the SDSS WDMS  binary sample  contains both
wide binaries, in which the white dwarf progenitor evolved as a single
star,  and PCEBs, in  which the  two stars  interacted when  the white
dwarf  progenitor evolved  off the  main sequence.   We  have obtained
radial  velocity information  spanning at  least two  nights  from our
follow-up   observations  of  several   hundred  SDSS   WDMS  binaries
\citep[e.g.][]{rebassa-mansergasetal08-1,            schreiberetal08-1,
  schreiberetal10-1} supplemented  by the SDSS  sub-spectra \citep[all
  SDSS spectra are  the result of combining at  least three individual
  sub-exposures,                      see][]{rebassa-mansergasetal07-1,
  rebassa-mansergasetal10-1}.   As   in  \citet{schreiberetal10-1}  we
classify  systems exhibiting  significant radial  velocity variations,
i.e.   the null  hypothesis that  radial velocity  is constant  can be
rejected on  a confidence  level $\geq0.9973$ ($3\,\sigma$),  and that
therefore  must have  short orbital  periods  ($\la$ a  few days),  as
PCEBs.  WDMS  binaries that do not exhibit  radial velocity variations
are designated  as strong  wide binary candidates,  and may  be either
genuine wide binaries,  or PCEBs that were not  recognized as such due
to (a combination of) a  low orbital inclination, long orbital period,
and  unlucky   phase  coverage.   We  discuss  the   number  of  PCEBs
potentially missed using these criteria in Sect.\,\ref{s-closebin}.

To  the 670  systems  analyzed by  \citet{schreiberetal10-1} we  added
seven systems  with uncertain secondary  mass and excluded  46 systems
with  rather uncertain  white  dwarf parameters.   From the  resulting
sample  of 629  systems  we considered  then  only objects  containing
hydrogen-rich (DA) white  dwarfs with effective temperatures exceeding
$12000$\,K.   Below this  limit  an increase  in  surface gravity  (as
obtained from spectral model fitting) is observed, probably related to
the description  of convection in  the framework of  the mixing-length
approximation \citep{koesteretal09-1,  tremblayetal10-1}.  Finally, we
considered  only WDMS  binaries with  errors in  the white  dwarf mass
smaller than  $0.1\,\Msun$.  The latter condition  is rather arbitrary
but the conclusions of the paper are independent on the exact value of
uncertainty we adopt.  The final  numbers of PCEB and wide WDMS binary
candidates in our  sample are 76 and 135  respectively (211 in total).
The  WDMS binaries  and their  white dwarf  masses  plus corresponding
errors  are   given  in  Table\,\ref{t-mass}.    The  radial  velocity
measurements for  the seven  objects that have  not been  presented in
\citet{schreiberetal10-1} but have been used in this work are given in
Table\,\ref{t-newrvs},  and  a  description  of the  observations  and
instrument  setup  can  be  found  in  Nebot  G\'omez-Mor\'an  et  al.
(2011b), in prep.

\begin{table}
\centering
\caption{\label{t-newrvs}  Radial velocities  and  heliocentric Julian
  dates (HJD) for the seven additional WDMS binaries used in this work
  in addition to the systems from \citet{schreiberetal10-1}.  The last
  column  indicates the telescope  where the  data were  taken: Gemini
  South (GS),  Very Large Telescope (VLT),  William Herschel Telescope
  (WHT).  If the radial velocities  were measured from SDSS spectra we
  use the quotation SDSS.}  \setlength{\tabcolsep}{1.0ex}
\begin{small}
\begin{tabular}{ccccc}
\hline
\hline
Object (SDSSJ) & HJD & RV  & error & obs.  \\
               &     &[$\kms$] & [$\kms$] & \\
\hline
  103837.22+015058.4   & 2454854.8083 &  21.6 &  11.5 & GS   \\  
                       & 2454855.8003 &  -5.5 &  12.6 & GS   \\  
                       & 2452317.7990 &   4.8 &  11.4 & SDSS \\  
  122752.72--015053.0  & 2454935.6564 & 113.6 &   5.8 & VLT  \\  
                       & 2454944.6149 &  97.7 &   6.3 & VLT  \\  
                       & 2451993.8006 & 120.1 &  14.0 & SDSS \\  
  135228.14+091039.0   & 2454514.8310 &   5.0 &   4.6 & GS   \\  
                       & 2454536.8304 &  58.5 &   9.8 & GS   \\  
                       & 2453559.1615 &  17.8 &  16.4 & SDSS \\  
  143551.64+043209.9   & 2454884.8206 &  -2.6 &   4.5 & VLT  \\  
                       & 2454892.8258 &   4.4 &   5.1 & VLT  \\  
                       & 2452023.8680 &  19.1 &  16.6 & SDSS \\  
  184117.99+410628.3   & 2454599.9360 &  20.1 &  15.2 & SDSS \\  
                       & 2454617.8310 &  -2.3 &  14.9 & SDSS \\  
                       & 2454617.8520 & -31.5 &  15.3 & SDSS \\  
  210426.91+101813.7   & 2454741.5720 & -90.5 &   5.9 & VLT  \\  
                       & 2454756.5199 &-103.3 &   9.2 & VLT  \\  
  223530.61+142855.0   & 2453919.6404 &-165.5 &   6.5 & WHT  \\  
                       & 2453920.5956 & 211.3 &   6.6 & WHT  \\  
\hline
\end{tabular}
\end{small}
\end{table}

\begin{figure*}
\centering
\includegraphics[angle=270,width=13cm]{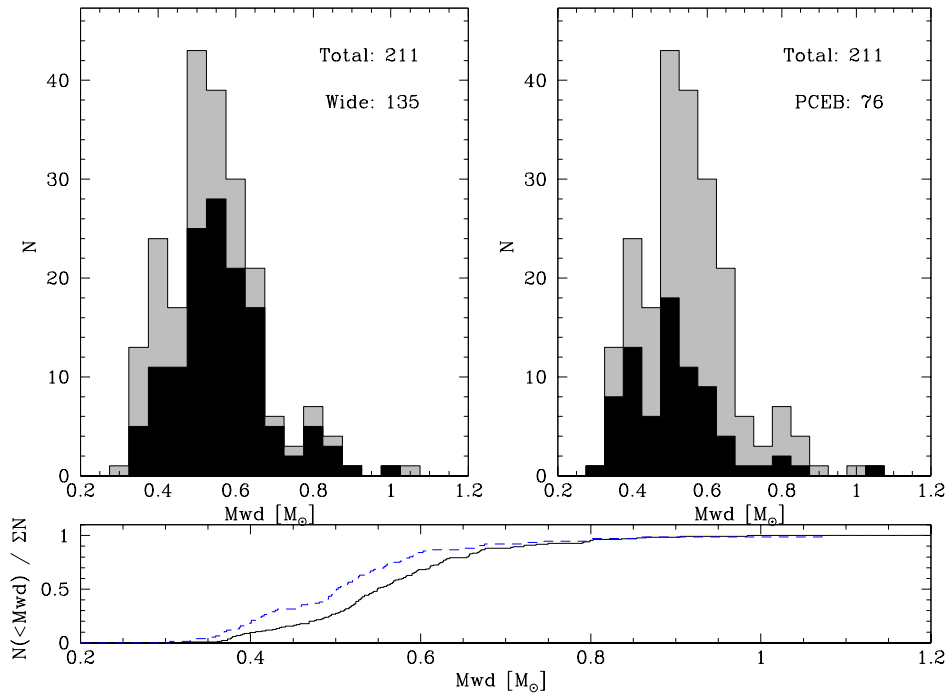}
\caption{Top panels: white dwarf  mass distributions obtained from the
  complete  sample of  WDMS  binaries (gray),  from  the PCEBs  (right
  panel, black), and from the wide WDMS binary candidates (left panel,
  black)  used  in  this  work.   Bottom panel:  the  cumulative  mass
  distributions of our  PCEBs (dashed blue line) and  wide WDMS binary
  candidates (black  line).  A two-sample  KS-test (Table\,\ref{t-ks})
  shows  that   the  two  distributions   represent  different  parent
  populations, i.e. are statistically independent.}
\label{f-wdmassdist}
\end{figure*}

\section{White dwarf mass distributions}
\label{s-result}

The white  dwarf mass distributions  of the complete WDMS  sample, the
PCEBs, and the  wide binary candidates are shown in  the top panels of
Fig.\,\ref{f-wdmassdist}. The mass distribution of the complete sample
is  bi-modal with  the strongest  peak near  $\sim0.55$\,\Msun,  and a
lower peak  near $\sim0.4$\,\Msun. The mass distribution  of the PCEBs
(right top panel) reveals a clear concentration of systems towards the
low-mass end  of the  distribution.  The number  of PCEB  white dwarfs
smoothly  declines  for  $M\ga0.55$\,\Msun.   In  contrast,  the  mass
distribution of the wide WDMS binary candidates (top left panel) shows
a  clear peak  at  $\sim0.55\,\Msun$, with  a  low-mass shoulder  near
$\sim0.4$\,\Msun.   Taking the  morphology of  the histograms  at face
value  clearly  suggests that  the  white  dwarfs  in PCEBs  follow  a
different mass distribution than  those in the wide binary candidates,
with a  substantial higher fraction  of low-mass white  dwarfs amongst
the PCEBs.

\begin{figure*}
\centering
\includegraphics[angle=-90,width=13cm]{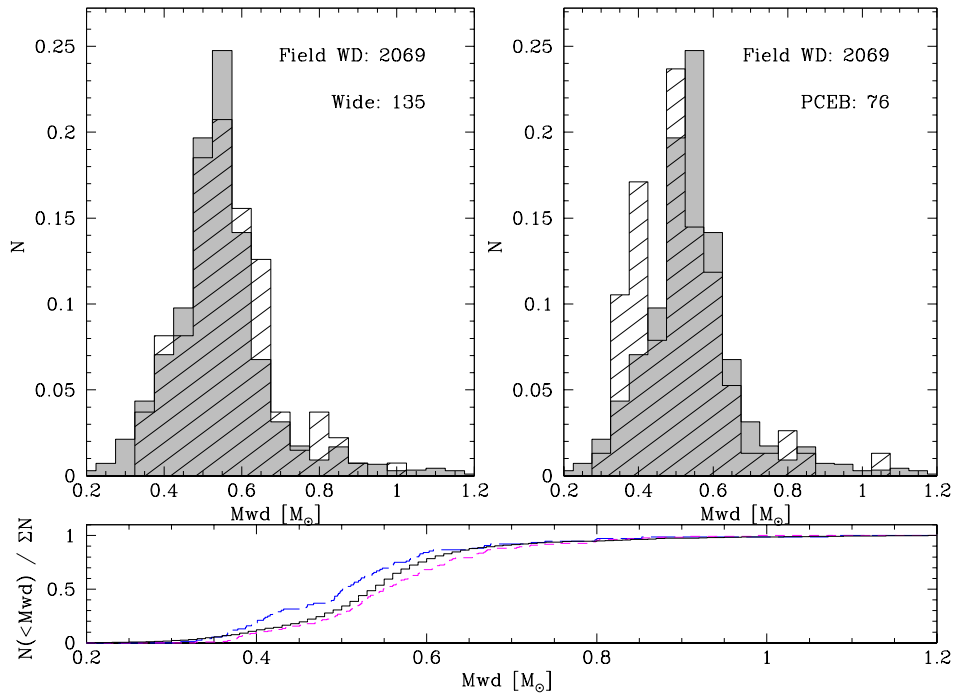}
\caption{Top panels: white dwarf  mass distributions obtained from the
  complete  sample  of  field   white  dwarfs  (gray  non-shaded,  see
  Table\,\ref{t-ks}  and Sect.\,\ref{s-comp}),  from the  PCEBs (right
  panel, white shaded), and from the wide WDMS binary candidates (left
  panel,  white  shaded)  used   in  this  work.   To  facilitate  the
  comparison  the distributions have  been normalized.   Bottom panel:
  the cumulative  mass distributions of our  PCEBs (blue long-dashed),
  wide WDMS binary candidates  (magenta short-dashed), and field white
  dwarfs (black).  KS-tests applied  among the three distributions are
  given in Table\,\ref{t-ks}.}
\label{f-wdmassdist2}
\end{figure*}

The  cumulative  mass  distributions  of  the PCEBs  and  wide  binary
candidates  (Fig.\,\ref{f-wdmassdist},   bottom  panel)  confirms  the
excess  of  low-mass  white  dwarfs  among the  PCEBs.   A  two-sample
Kolmogorov-Smirnov (KS) test on  the two cumulative mass distributions
rejects  the  null-hypothesis   with  $99.5859$  per  cent  confidence
(Table\,\ref{t-ks}).   In  other  words,  the  two  distributions  are
statistically independent and the  accumulation of PCEBs at low masses
is highly significant.
 
The  high fraction  of  low-mass white  dwarfs  among the  PCEBs is  a
fundamental observational  confirmation of the  hypothesis that common
envelope evolution will often lead  to a truncation of the core growth
in the more massive binary companion.

\section{Comparing SDSS single white dwarf and WDMS binary mass distributions}
\label{s-comp}

The mass distribution  of field white dwarfs has  been subject to more
than  three decades of  investigations. Early  investigations revealed
the   strong    clustering   of   white   dwarf    masses   close   to
0.6\,\Msun\   \citep[e.g.][]{koesteretal79-1}.    Towards  the   early
nineties,  white dwarf samples  with high-quality  spectroscopy became
sufficiently  large  to  reveal  additional structure  in  their  mass
distributions, in particular  a low-mass component comprising $\sim10$
per cent of the  total white dwarf population \citep{bergeronetal92-1,
  bragagliaetal95-1},  confirming   predictions  of  binary  evolution
theory  \citep{iben+tutokov86-2, iben90-1}.   The  estimated published
fractions of low-mass white dwarfs have remained broadly constant over
time \citep{liebertetal05-1, kepleretal07-1}.

In the previous section, we  have shown that the mass distributions of
white dwarfs in PCEBs and wide binary candidates differ significantly,
with  a larger  fraction of  low-mass  white dwarfs  among the  PCEBs.
Here, we  compare the white dwarf  mass distribution of  the PCEBs and
the wide WDMS binary candidates with that of field white dwarfs.

When  comparing  white  dwarf  mass distributions,  many  factors  can
introduce  systematic  differences,   such  as  the  wavelength  range
\citep{kepleretal06-1}         and        signal-to-noise        ratio
\citep{liebertetal05-1} of the observational  data, the details of the
fitting  procedures, and  the  model grids  adopted  for the  analysis
\citep[e.g.][]{tremblayetal10-1}. To  keep such systematic differences
to  a minimum, we  decided to  fit a  large sample  of field  DA white
dwarfs   from   SDSS   with   the   same  routine   and   model   grid
\citep{rebassa-mansergasetal07-1,  koesteretal05-1}  as  used for  the
WDMS  binaries. We  adopted the  DA sample  of \citet{kepleretal07-1},
which consists  of white dwarfs classified as  single and non-magnetic
\citep[for  details   see][]{eisensteinetal06-1},  but  retrieved  the
corresponding DR\,7 spectra which  were processed by the same pipeline
as the WDMS binary spectra. For the vast majority of the 7167 spectra,
our  results  were consistent  with  those of  \citet{kepleretal07-1}.
Significant differences were  found for 265 spectra, where  it is most
likely that either our fits, or those of Kepler et al. chose the wrong
of      the     ``hot''      and      ``cold''     solutions      (see
\citealt{rebassa-mansergasetal07-1}   for   a   discussion   of   this
problem). These 265 systems were removed from the field DA sample.

Fig.\,\ref{f-wdmassdist2} (top  left panel) illustrates  that the mass
distributions of  field white  dwarfs and those  in wide  binaries are
very similar,  suggesting that white dwarfs in  both subsamples evolve
in a very similar way.  A two-sample KS test of the field white dwarfs
vs.   wide WDMS  binary candidates,  limited  to 2069  systems with  a
0.1\,\Msun\,  uncertainty and effective  temperatures $>  12000\,$K in
the white dwarf parameters, gives a KS probability of $\sim$5 per cent
(see     Table\,\ref{t-ks}     and     the     bottom     panel     of
Fig.\,\ref{f-wdmassdist2}).  There  are, hence, no  strong indications
for white  dwarfs in  wide WDMS binaries  evolving in a  different way
from isolated field white dwarfs.   We note that the relatively low KS
probability  obtained  in  this  exercise  may  be  a  consequence  of
selection  effects in SDSS  affecting the  mass distributions,  and of
some PCEBs not  identified by our radial velocity  survey hiding among
the wide  WDMS binary  candidates.  Finally, it  is worth  noting that
extremely low-mass ($\la0.35\Msun$) white dwarfs seem to be present in
the apparently  single white  dwarf sample but  absent among  the wide
WDMS binaries.   Further discussions of  these issues are  provided in
the following Sections.

A KS test between the PCEB and field white dwarf mass subsamples gives
a KS probability of only  0.77 per cent (see Table\,\ref{t-ks} and the
bottom  panel of  Fig.\,\ref{f-wdmassdist2}), clearly  confirming that
white  dwarfs in PCEBs  have indeed  evolved in  a different  way than
field white dwarfs.

\begin{table}
\centering
\caption{\label{t-ks}  Results of two-sample  Kolmogorov-Smirnov tests
  comparing  the cumulative  white  dwarf mass  distributions of  four
  different samples used in  this work: ``PCEB'', ``wide WDMS'', ``all
  WDMS''  and  ``field WD'',  standing  for  PCEBs,  wide WDMS  binary
  candidates,  the complete  WDMS binary  sample, and  field  DA white
  dwarfs     from     \citet{kepleretal07-1}     respectively     (see
  Sect.\,\ref{s-data} and Sect.\,\ref{s-comp} for details). The number
  of  systems for each  sample are  also indicated  in the  second and
  fourth columns.}  \setlength{\tabcolsep}{2ex}
\begin{tabular}{ccccc}
\hline
\hline
Sample\,1 & N & Sample\,2 & N & KS Prob.   \\
          &   &           &   & [per cent] \\
\hline
PCEB     & 76   & wide WDMS & 135 & 0.414 \\
field WD & 2069 & wide WDMS & 135 & 4.731 \\
field WD & 2069 & PCEB      &  76 & 0.770 \\
\hline
\end{tabular}
\end{table}

\section{PCEB fractions}
\label{s-closebin}

We  have shown  that the  mass distributions  of PCEBs  and  wide WDMS
binaries  differ  significantly.  While  the  former  contains a  much
larger fraction  of low-mass  white dwarfs, the  latter is  similar to
single white  dwarf mass distributions.  In other  words, the fraction
of PCEBs among WDMS binaries containing low-mass white dwarfs seems to
be significantly  larger than  for WDMS binaries  containing high-mass
white dwarfs.   However, caution is needed to  interpret these results
as  most of our  PCEBs have  been classified  according to  few radial
velocity  measurements spread  over at  least two  nights.   This PCEB
identification method  is obviously less  sensitive to PCEBs  with low
inclinations and/or long orbital  periods and hence a certain fraction
of PCEBs will be hiding among the wide WDMS binary candidates.  In the
following we correct for this bias using Monte-Carlo methods.

\begin{figure}
\centering
\includegraphics[angle=-90,width=\columnwidth]{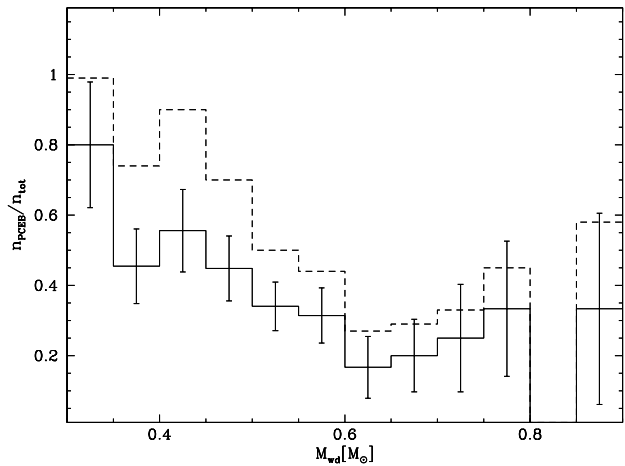}
\caption{The  observed  (solid  line)  and  the  detection-probability
  corrected (dotted  line) fraction of PCEBs among  SDSS WDMS binaries
  as  a function  of white  dwarf  mass.  Virtually  all systems  with
  $\Mwd\leq0.35\,\Msun$ and  $\sim80$ per  cent of systems  with white
  dwarfs  in the  mass  range $\Mwd=0.35-0.5\,\Msun$  are PCEBs.   For
  high-mass white dwarfs ($\Mwd>0.5\,\Msun$) the fraction decreases to
  $\sim40$ per cent.}
\label{f-detec}
\end{figure}

To determine  the most likely fraction  of PCEBs for  each white dwarf
mass-bin  (we adopt  a  mass-bin of  0.05\Msun)  we need  to know  the
probability for each  system being a PCEB.  For  those systems showing
strong radial  velocity variations  this probability equals  one.  For
those systems  classified as  wide binaries we  need to  calculate the
PCEB  detection  probability.   Taking   into  account  the  times  of
observations   and  the  stellar   masses  we   performed  Monte-Carlo
simulations  similar to  those presented  in \citet{schreiberetal10-1}
and Nebot G\'omez-Mor\'an et al.  (2011b, in prep.), i.e.  for typical
PCEB   orbital  periods   ($2\,$hr$-2\,$days)  we   randomly  selected
$10\,000$  times the phases  and inclinations\footnote{The  phases are
  assumed  to  be  uniformely  distributed  over  $0-2\pi$  while  the
  inclinations  have  been  chosen  uniformly  from  the  distribution
  $\sin(i)$.} for each system  and calculated the corresponding radial
velocities.   Averaging the fraction  of $>3\,\sigma$  radial velocity
variations over  the orbital period gives the  averaged PCEB detection
probability  for each  system ($e_i$).   The most  likely  fraction of
PCEBs can then be  calculated analogous to \citet{maxtedetal00-3}.  We
assume two models,  i.e.  all WDMS binaries are PCEBs  (model M1) or a
certain fraction ($f_A$) of WDMS binaries are PCEBs (model M2).  Using
Bayes' theorem we  obtain for the probability ratio  of the two models
given our data $D$:
\begin{equation} 
\frac{P(D|M2)}{P(D|M1)}=\frac{f_A^{N_A} \prod q_i}{\prod p_i},
\end{equation}
where
\begin{equation}
p_i=1-e_i,
\end{equation}
and
\begin{equation}
q_i=f_A[1-e_i].
\end{equation}
The measured values  entering these equations are the  number of PCEBs
$N_A$  and  the  detection  probabilities  of  the  wide  WDMS  binary
candidates  ($e_i$).  We  have  calculated the  probability ratio  for
values of  $f_A$ ranging from  0 to 1  with a step-size of  $0.01$ for
each  white dwarf  mass-bin.   The maximum  value  indicates the  most
probable  fraction of  PCEBs  which is  shown  by the  dashed line  in
Fig.\,\ref{f-detec}  together with the  measured PCEB  fraction (solid
line).  It appears  that the bias-corrected PCEB fraction  is close to
$100$ per  cent for the  lowest mass systems and  decreases relatively
sharply at $\Mwd\sim0.5\,\Msun$ from $\sim\,80$ per cent to $\sim\,40$
per cent.

\section{Wide WDMS binaries containing low-mass white dwarfs}
\label{s-wideor}

\begin{table*}
\centering
\caption{\label{t-stats} PCEB and low-mass white dwarf fractions among
  our WDMS binaries  for different values of $sf$,  where $sf$ defines
  the  sample of  WDMS  binaries used  (see Sect.\,\ref{s-wideor}  for
  details).  Here, $f_{\mathrm{\Mwd>0.5}}$ and $f_{\mathrm{\Mwd<0.5}}$
  give the fraction  of WDMS binaries with white  dwarfs of $\Mwd>0.5$
  and    $\Mwd\leq0.5$    respectively.     $N_{\mathrm{PCEB,    h}}$,
  $N_{\mathrm{WIDE, h}}$, $N_{\mathrm{PCEB, l}}$ and $N_{\mathrm{WIDE,
      l}}$ are the  number of PCEBs and wide  WDMS binaries containing
  high-mass  (h)  and  low-mass  (l)  white dwarfs.   For  $sf  =  0$,
  $N_{\mathrm{PCEB, h}} +  N_{\mathrm{WIDE, h}} + N_{\mathrm{PCEB, l}}
  + N_{\mathrm{WIDE, l}} = 211$, our complete sample of WDMS binaries.
  Furthermore,  $f_{\mathrm{PCEB,  lm}}$  and $f_{\mathrm{PCEB,  hm}}$
  represent  the measured fraction  of PCEBs  among all  WDMS binaries
  containing  low-mass and  high-mass white  dwarfs  respectively, and
  $f_{\mathrm{PCEB,   l}}$   and   $f_{\mathrm{PCEB,  h}}$   are   the
  corresponding bias-corrected values.  Finally $f_{\mathrm{WIDE, m}}$
  and  $f_{\mathrm{WIDE, c}}$  are the  observed fraction  of low-mass
  white  dwarfs among wide  WDMS binaries  and the  corresponding bias
  corrected value.}  \setlength{\tabcolsep}{1.1ex}
\begin{small}
\begin{tabular}{ccccccccccccc}
\hline
\hline
$sf$ & $f_{\mathrm{\Mwd>0.5}}$ & $N_{\mathrm{PCEB, h}}$ & $N_{\mathrm{WIDE, h}}$ &
 $f_{\mathrm{PCEB, hm}}$ & $f_{\mathrm{PCEB, h}}$ &  
$f_{\mathrm{\Mwd<0.5}}$ & $N_{\mathrm{PCEB, l}}$ & $N_{\mathrm{WIDE, l}}$ &
$f_{\mathrm{PCEB, lm}}$ & $f_{\mathrm{PCEB, l}}$ &
$f_{\mathrm{WIDE, m}}$ & $f_{\mathrm{WIDE, c}}$ \\
\hline
$0$ & $0.65\pm0.03$ & $39$ & $99$ & $0.28\pm0.04$ & 
$0.41_{-0.05}^{+0.06}$ & $0.35\pm0.03$ & $37$ & $36$ & $0.51\pm0.06$ &
$0.80_{-0.09}^{+0.08}$ & $0.27\pm0.04$ & $0.15_{-0.01}^{+0.01}$ \\
$0.5$ & $0.67\pm0.04$ & $30$ & $85$ & $0.26\pm0.04$ & 
$0.38_{-0.05}^{+0.06}$ & $0.33\pm0.04$ & $29$ & $28$ & $0.51\pm0.07$ &
$0.80_{-0.10}^{+0.09}$ & $0.25\pm0.04$ & $0.13_{-0.01}^{+0.01}$ \\
$1$ & $0.68\pm0.04$ & $24$ & $72$ & $0.25\pm0.04$ & 
$0.36_{-0.06}^{+0.07}$ & $0.32\pm0.04$ & $24$ & $21$ & $0.53\pm0.07$ &
$0.85_{-0.11}^{+0.08}$ & $0.23\pm0.04$ & $0.10_{-0.01}^{+0.01}$ \\
$1.5$ & $0.70\pm0.04$ & $20$ & $54$ & $0.27\pm0.05$ & 
$0.39_{-0.07}^{+0.08}$ & $0.30\pm0.04$ & $17$ & $15$ & $0.53\pm0.09$ &
$0.84_{-0.13}^{+0.09}$ & $0.22\pm0.05$ & $0.10_{-0.02}^{+0.02}$ \\
$2$ & $0.68\pm0.05$ & $17$ & $37$ & $0.31\pm0.06$ & 
$0.45_{-0.08}^{+0.09}$ & $0.32\pm0.05$ & $16$ & $9$ & $0.64\pm0.10$ & 
$0.99_{-0.14}^{+0.01}$ & $0.20\pm0.06$ & $0.01_{-0.03}^{+0.03}$ \\
\hline
\end{tabular}
\end{small}
\end{table*}

\begin{figure}
\centering
\includegraphics[angle=-90,width=\columnwidth]{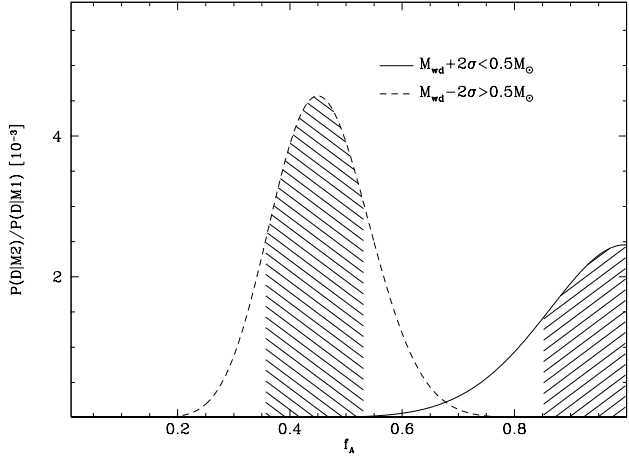}
\caption{Probability ratio  as a function  of PCEB fraction  $f_A$ for
  WDMS binaries containing low-mass (solid line) and high-mass (dashed
  line) white dwarfs.   The most likely fraction of  PCEBs is given by
  the highest  probability ratio, i.e.  $f_A=0.99$  for low-mass white
  dwarfs, and $f_A=0.45$ for high-mass  white dwarfs. The error of the
  most likely value is calculated by requiring the probability on each
  side  of the  maximum to  reach $68.27$  per cent  indicated  by the
  shaded regions.}
\label{f-prat}
\end{figure}

Because of their uncertain origin,  the fraction of wide WDMS binaries
containing  low-mass  white  dwarfs  is  of  crucial  importance.  The
measured fraction  of systems containing a low-mass  white dwarf among
those  WDMS binaries  that  do not  show  significant radial  velocity
variations  is $27$  per  cent.  However, as  shown  above, a  certain
fraction of these systems are certainly PCEBs that we did not identify
due to low inclinations, long orbital periods, or unfortunate sampling
of the  orbital phase (see Sect.\,\ref{s-closebin}).   In this section
we  investigate in  more detail  the true  fraction of  low-mass white
dwarfs among wide WDMS binaries implied by our observations.  For this
purpose we first divide  our sample in high-mass ($\Mwd>0.5\Msun$) and
low-mass ($\Mwd\leq0.5\Msun$) systems and determine the bias corrected
fraction  of PCEBs  in both  sub-samples.  As  the uncertainty  of our
white  dwarf mass  estimates is  typically $\sigma\sim0.05-0.1\,\Msun$
and therefore the theoretically predicted clear separation of low-mass
and  high-mass white dwarfs  at $\Mwd=0.5\,\Msun$  is smeared  out, we
additionally      require      $\Mwd+sf\times\sigma<0.5\Msun$      and
$\Mwd-sf\times\sigma>0.5\Msun$ with $sf=0, 0.5, 1.0, 1.5, 2.0$. $sf=0$
corresponds to  our complete WDMS  binary sample.  As  $sf$ increases,
the number  of WDMS  binaries that we  consider drops, because  we are
excluding systems close to the  boundary separating low and high white
dwarf  masses.  As  in Sect.\,\ref{s-closebin},  the most  likely PCEB
fraction  is  given by  the  maximum  value  of $P(D|M2)/P(D|M1)$.  An
example of  the probability ratio  $P(D|M2)/P(D|M1)$ as a  function of
the assumed PCEB fractions  $f_A$ is shown in Fig.\,\ref{f-prat}.  For
systems  containing  low-mass  white  dwarfs the  maximum  probability
corresponds to $f_A=f_{\mathrm{PCEB,l}}=0.99$  (see the solid curve in
Fig.\,\ref{f-prat})  while  the most  likely  PCEB  fraction for  WDMS
binaries containing  high-mass white dwarfs  is significantly smaller,
i.e.    $f_A=f_{\mathrm{PCEB,h}}=0.45$  (see   the  dashed   curve  in
Fig.\,\ref{f-prat}).  The error of the most likely value is calculated
by  requiring the probability  on each  side of  the maximum  to reach
$68.27$   per  cent,   as   indicated  by   the   shaded  regions   in
Fig.\,\ref{f-prat}.  This results  in asymmetric errors reflecting the
asymmetry of the probability functions.

Finally,  the  fraction  of  low-mass  white dwarfs  among  wide  WDMS
binaries  ($f_{\mathrm{WIDE,   c}}$)  can  be   calculated  using  the
estimated      PCEB      fractions      $f_{\mathrm{PCEB,h}}$      and
$f_{\mathrm{PCEB,l}}$ according to
\begin{equation}
\label{eq-wide}
f_{\mathrm{WIDE, c}}=\frac{N_{\mathrm{WDMS,l}}(1-f_{PCEB,l})}{N_{\mathrm{WDMS,l}}(1-f_{PCEB,l})+N_{\mathrm{WDMS,h}}(1-f_{PCEB,h})},
\end{equation}
where  $N_{\mathrm{WDMS,h}}$ and  $N_{\mathrm{WDMS,l}}$  represent the
number  of  WDMS  binaries  containing high-mass  and  low-mass  white
dwarfs, respectively,  in the  considered sample.  The  resulting PCEB
fractions and  the estimated fraction  of low-mass white  dwarfs among
wide  systems are  listed  in Table  \ref{t-stats}  for the  different
choices of $sf$.  We would like to highlight two numbers given in this
table.  On the one hand we see that the fraction of wide WDMS binaries
containing   low-mass  white   dwarfs   ($f_{\mathrm{WIDE,  c}}$)   is
$\sim1-15$ per  cent.  The origin of  these systems is  unclear and is
further discussed  in Sect.\,\ref{s-discuss}.  On the  other hand, the
most likely PCEB fraction  for WDMS binaries containing low-mass white
dwarfs  ($f_{\mathrm{PCEB,l}}$) is $\sim80-99$  per cent.   This shows
that  the large majority  of WDMS  binaries containing  low-mass white
dwarfs have formed as a consequence of common envelope evolution.  For
comparison, the most likely PCEB fraction for WDMS binaries containing
high-mass      white     dwarfs     is      significantly     smaller,
i.e. $f_{\mathrm{PCEB,h}}  \sim 41-45$ per cent.  In  the next section
we  estimate the  fraction of  PCEBs among  {\em{all}}  low-mass white
dwarfs (WDMS binaries with low mass white dwarf plus apparently single
low-mass white dwarfs).

\section{Are most low-mass white dwarfs formed in close binaries?}
\label{s-are}

We  have  shown  that  $\sim80-99$  per  cent  of  the  WDMS  binaries
containing low-mass  white dwarfs  are PCEBs. Roughly  10 per  cent of
apparently single white  dwarfs also have low masses.   If we assume a
binary fraction of  60 per cent for solar type  and more massive stars
\citep[e.g.][]{duquennoy+mayor91-1} and take  the fraction of low-mass
white dwarfs in  our sample of WDMS binaries  ($f_{\Mwd<0.5}$) at face
value,  the 80-99  per cent  of PCEBs  among WDMS  binaries containing
low-mass white dwarfs make up a fraction of
\begin{equation} 
\label{eq-all}
    \frac{0.6\times\,f_{\mathrm{\Mwd<0.5}}\times\,f_{\mathrm{PCEB,l}}}
{(0.6\times\,f_{\Mwd<0.5}+0.4\times\,0.1)}\simeq\,0.67-0.82
\end{equation} 
of all  low-mass white dwarfs (those  in WDMS binaries  in addition to
those low-mass white dwarfs that are apparently single).  In contrast,
an  analogous estimate  for  high-mass white  dwarfs  gives that  only
$\sim21-24$ per cent of {\em{all}} high-mass white dwarfs (those among
WDMS binaries plus those apparently single high-mass white dwarfs) are
formed in close binaries.

\section{Potential selection effects in the white dwarf mass distributions}
\label{s-seleff}

Taking into  account observational biases  due to our  radial velocity
method of identifying  PCEBs in the white dwarf  mass distributions of
close and wide  WDMS binaries, we calculated in  the previous sections
the  most likely fraction  of wide  WDMS binaries  containing low-mass
white  dwarfs  (Sect.\,\ref{s-wideor}),   as  well  as  estimated  the
fraction  of  all low-mass  white  dwarfs  that  has formed  in  close
binaries (Sect.\,\ref{s-are}). However, selection effects intrinsic to
the  SDSS  survey  may  also   have  affected  our  white  dwarf  mass
distributions.  In  what follows we  consider the implications  of the
SDSS  magnitude limits and  the identification  of WDMS  binaries from
their SDSS spectra.

\begin{figure*}
\includegraphics[width=0.68\columnwidth]{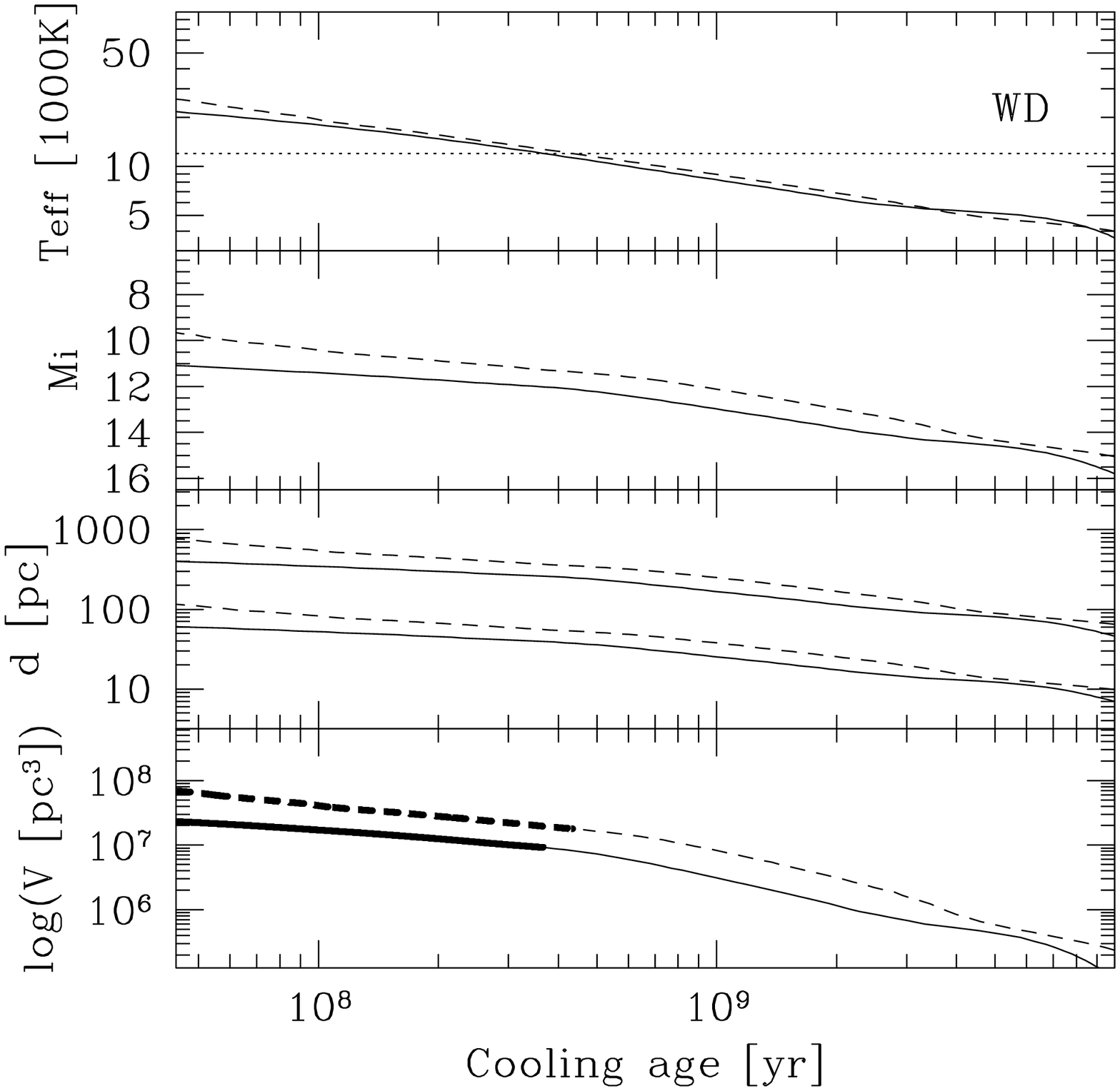}
\hfill
\includegraphics[width=0.68\columnwidth]{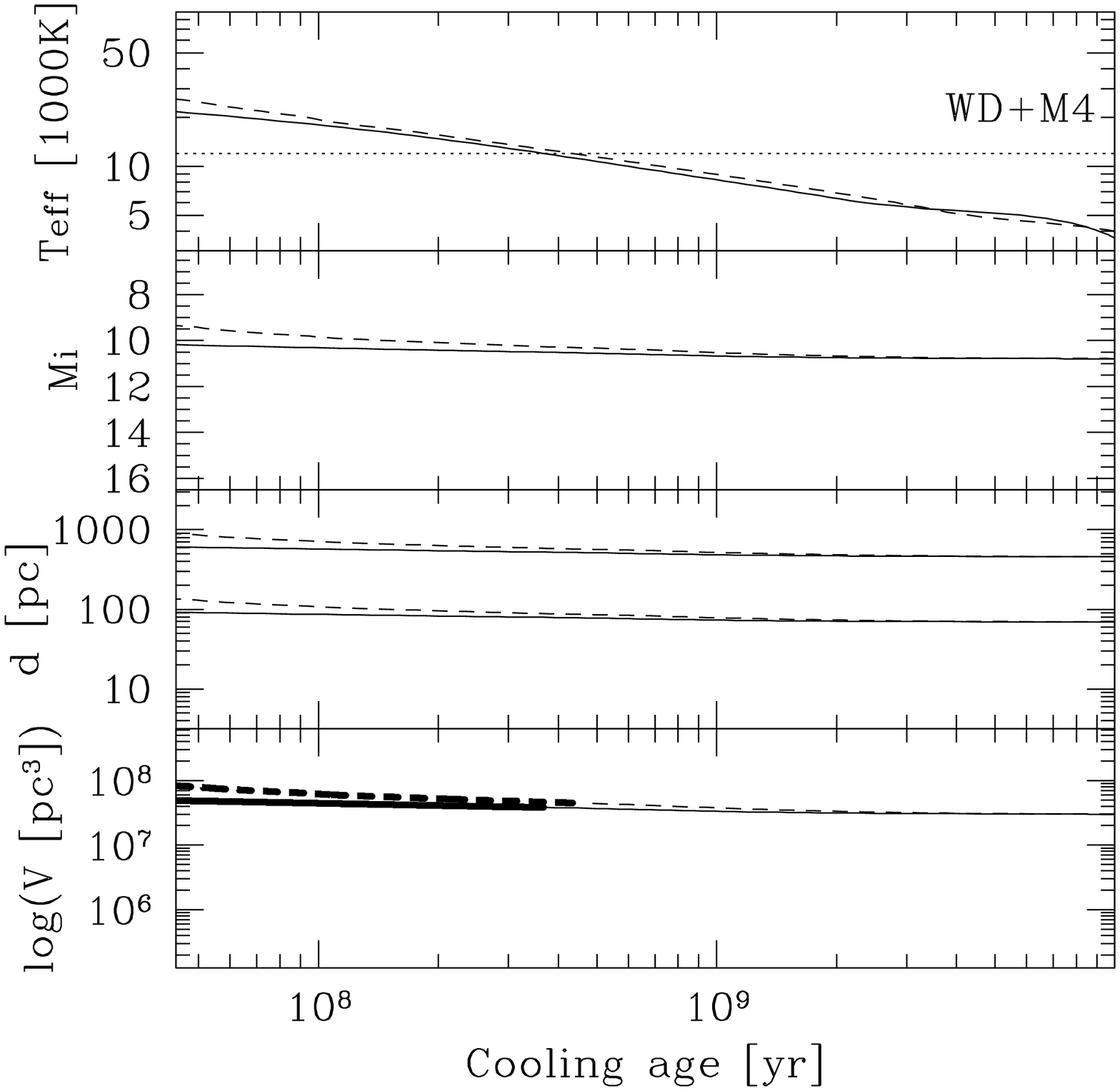}
\hfill
\includegraphics[width=0.68\columnwidth]{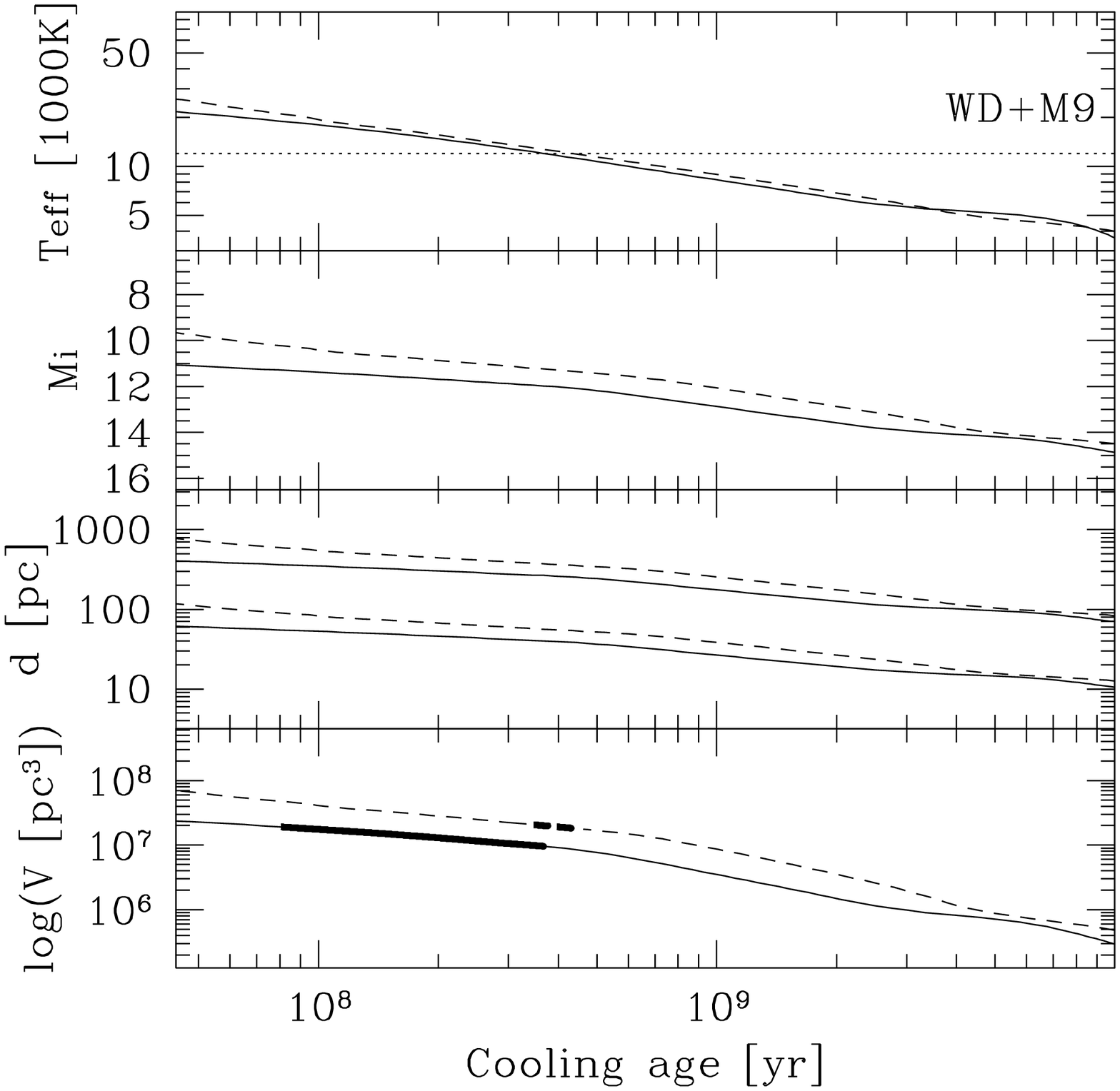}
\caption{The  effective survey  volume as  a function  of  white dwarf
  cooling age for single white dwarfs (left), a WDMS binary with an M4
  companion (middle) and  a WDMS binary with an  M9 companion (right).
  From top to  bottom, the panels show: (1)  the effective temperature
  for  a 0.6\,\Msun\ CO-core  (solid line)  and a  0.4\,\Msun\ He-core
  white  dwarf (dashed line).   The effective  temperature $>12000$\,K
  cut-off  of  our WDMS  binary  sample  is  indicated by  the  dotted
  horizontal  line.   (2)  the  absolute  $i$-band  magnitude  of  the
  system. (3)  the saturation limit  of SDSS ($\simeq15.5$)  implies a
  minimum distance  for the systems  (lower two curves),  the limiting
  magnitude of the main quasar survey ($\simeq19.1$) translates into a
  maximum distance (upper two curves). (4) the effective survey volume
  for an assumed scale  height $H_\mathrm{z}=200$\,pc and a Sloan-type
  sky coverage. The relative bias for He-core white dwarfs relative to
  CO-core   white   dwarfs   (Fig.\,\ref{f-bias_summary})   has   been
  calculated  by  integrating the  effective  survey  volume over  the
  cooling age  range in  which the systems  can be identified  as WDMS
  binaries from the SDSS spectroscopy  (solid and dashed bold lines in
  the bottom middle and bottom right panels).  For single white dwarfs
  the  integral  is  restricted   to  cooling  ages  corresponding  to
  $\Teff(\mathrm{min})=12000$\,K,  and  $\Teff(\mathrm{max})$  to  the
  highest  values available in  the cooling  models (solid  and dashed
  bold lines in the bottom left panel).}
\label{f-bias_volume}
\end{figure*}

\subsection{The impact of a magnitude limited sample}

A recurrent problem in using large-area surveys for population studies
is   that   they   are   \textit{magnitude   limited},   rather   than
\textit{volume  limited},  and  hence  the  survey volume  will  be  a
function  of the intrinsic  brightness, or  absolute magnitude  of the
stars in the  sample under study.  In the case  of SDSS, the magnitude
limits are $\simeq15.5$ across all  bands at the bright end, where the
imaging  saturates,   and  $i\simeq19.1$  at  the   faint  end,  which
corresponds to the  limit of the main quasar  survey, which provided a
large fraction of  the SDSS WDMS binaries, but  is also representative
as a generic limit for  the bulk of the spectroscopic follow-up within
SDSS.

The  absolute magnitude  of white  dwarfs depends  on  their effective
temperature and their radius, with the temperature decreasing with age
as the white dwarfs cool, and the radius is decreasing with increasing
mass, as  they follow a  mass-radius relation for  degenerate objects.
We have adopted the cooling sequences of \citet{benvenuto+althaus99-1}
and \citet{bergeronetal95-2}\footnote{We used Bergeron's updated grids
  available                                                          at
  http://www.astro.umontreal.ca/$\sim$bergeron/CoolingModels/}       to
compute the effective temperature  and absolute $i$-band magnitude for
a  0.4\,\Msun\ He-core  white dwarf  and a  0.6\,\Msun\  CO-core white
dwarf (Fig.\,\ref{f-bias_volume}, left, top two panels).  Adopting the
SDSS  magnitude   limits  mentioned  above,   the  absolute  magnitude
translates  into a  minimum and  maximum  distance at  which SDSS  may
obtain follow-up spectroscopy for the star (Fig.\,\ref{f-bias_volume},
left, third panel from the top).  The \textit{effective} survey volume
of SDSS  is calculated  by integrating, for  all cooling  ages $\tau$,
over a spherical cap in galactic coordinates for $b>30$ and an adopted
scale  height of $H_z=200$\,pc  for distances  $d_\mathrm{min}\le d\le
d_\mathrm{max}$      (Fig.\,\ref{f-bias_volume},      left,     bottom
panel)\footnote{A more  detailed approach  would require to  take into
  account  the exact  tiling of  the SDSS  spectroscopy,  however this
  would not significantly change the results.}.  For any given $\tau$,
the     ratio    of     the    two     effective     survey    volumes
$V_\mathrm{He}/V_\mathrm{CO}$ gives the probability ratio of finding a
He-core white dwarf  compared to a CO-core white  dwarf. To obtain the
total  probability ratio,  we need  to  take into  account that  white
dwarfs of different mass and core composition cool at different rates.
Assuming a constant  formation rate of white dwarfs,  we integrate the
survey volume over the cooling age,

\begin{equation}
\int_{\Teff(\mathrm{min})}^{\Teff(\mathrm{max})} Vd\tau
\label{e-vdt}
\end{equation}

We   restrict  this   integral  to   cooling  ages   corresponding  to
$\Teff(\mathrm{min})=12000$\,K, the temperature cut-off imposed on our
sample by the ``Balmer line problem'' outlined in Sect.\,\ref{s-data},
and  $\Teff(\mathrm{max})$  to the  highest  values  available in  the
cooling  models  (corresponding to  such  low  cooling  ages that  the
integral is  a good approximation  of the total age-volume  space). We
find that, for field white  dwarfs, it is $\sim2.75$ times more likely
to   find  a  0.4\,\Msun\   He-core  white   dwarf  within   the  SDSS
spectroscopic follow-up than a 0.6\,\Msun\ CO-core white dwarf.

\subsection{Identifying the spectroscopic composite nature of a WDMS binary}

In order to identify a  WDMS binary from SDSS spectroscopy, both stars
need to  be detectable within  the composite spectrum.   For late-type
secondary  stars, this  implies  an  upper limit  on  the white  dwarf
temperature,  $\Teff(\mathrm{max})$,  at  which  we will  be  able  to
discern the companion in the SDSS spectrum.  Conversely, the detection
of white dwarfs next to early-type companions results in a lower limit
on   the  white   dwarf  temperature,   $\Teff(\mathrm{min})$.   These
temperature limits will depend somewhat on the white dwarf radius, and
hence its mass.  We have  simulated composite spectra for a wide range
of  white dwarf  temperatures and  companion spectral  types, adopting
either a 0.4\,\Msun\ He-core or a 0.6\,\Msun\ CO-core white dwarf, and
subjected  those spectra to  the same  identification criteria  as the
observed SDSS spectra \citep{rebassa-mansergasetal10-1}.  Full details
of  this  analysis will  be  given in  a  forthcoming  paper where  we
investigate  the   overall  completeness  of  the   SDSS  WDMS  binary
population.  Here,  we restrict the approach to  estimate the relative
bias  within that  WDMS binary  population with  respect to  the white
dwarf mass.   For the latest spectral  type companions, M8  and M9, we
find   $\Teff(\mathrm{max})=16\,000$\,K   and   13\,000\,K   for   the
0.4\,\Msun\ He-core  white dwarf, and $\Teff(\mathrm{max})=24\,000$\,K
and 19\,000\,K for the 0.6\,\Msun\ CO-core white dwarf.  For all other
companion  spectral  types,  we fix  $\Teff(\mathrm{max})=26\,000$\,K,
limited      by      the      He-core      cooling      models      of
\citet{benvenuto+althaus99-1}.   The   corresponding  cooling  age  is
sufficiently  small   that  the  integrals  defined   above  are  well
approximated.  The lower  limit of the white dwarf  temperature is set
to  $\Teff(\mathrm{min})=12\,000$\,K  (Sect.\,\ref{s-data}), with  the
only exception being that the  0.6\,\Msun\ CO-core white dwarf next to
M0 companions needs $\Teff(\mathrm{min})=14\,000$\,K.

\begin{figure}
\includegraphics[angle=-90,width=\columnwidth]{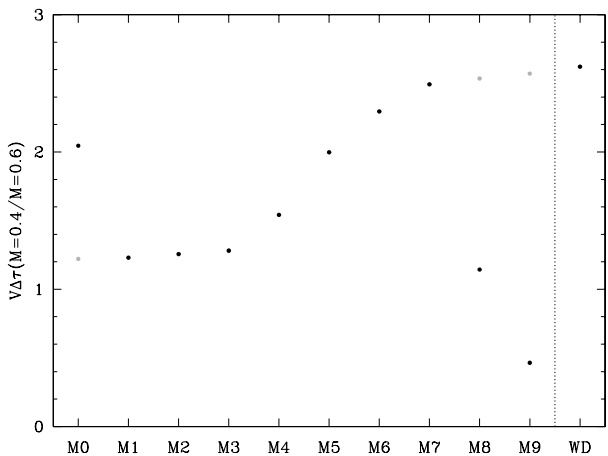}
\caption{Relative   probability  ratio  of   finding  a   WDMS  binary
  containing a He-core  white dwarf rather than a  CO-core white dwarf
  among objects  with SDSS spectroscopy  obtained as part of  the main
  quasar survey, with the spectral  type of the companion star running
  from  M0  (left)  to  M9  (right).  The  right-most  dot  gives  the
  probability ratio for single white  dwarfs. The black dots take into
  account both the effective survey volume (Fig.\,\ref{f-bias_volume})
  as well  as the  spectral-type dependent limits  on the  white dwarf
  temperature  (and  hence  cooling  age)  imposed  by  the  need  for
  detecting signatures  of both stars  in the composite  SDSS spectra.
  Shown  as gray  dots  are  the the  probability  ratios taking  into
  account only the effective survey volumes.}
\label{f-bias_summary}
\end{figure}

We then  repeat the calculation  carried out in the  previous Section,
adding  the absolute $i$-band  magnitude of  the companion  (using the
values  of \citealt{westetal05-1})  to that  of the  white  dwarf, and
integrating the  survey volume between the  cooling ages corresponding
to the  lower and  upper limits on  the white dwarf  temperature.  The
evolution of the $i$-band  magnitude $M_i$, the distance limits within
SDSS, and the  survey volume as a function of  white dwarf cooling age
are given  in Fig.\,\ref{f-bias_volume} for M4 (middle  panels) and M9
(right panels)  companions.  For the M4 companion,  $M_i$ converges to
the absolute magnitude of the secondary star as the white dwarf cools,
resulting in only  mild variations in the survey  volume as a function
of cooling age.  In contrast, the M9 companion contributes only little
to the  $i$-band even late in  the evolution of the  binary, and hence
the change  in the survey volume is  very similar to that  of a single
white  dwarf.  Finally,  we calculate  the  ratio of  the integral  in
Eq.(\ref{e-vdt}) for the 0.4\,\Msun\  He-core white dwarf with respect
to  the 0.6\,\Msun\ CO-core  white dwarf  for each  companion spectral
type (Fig.\,\ref{f-bias_summary}).   This ratio can  be interpreted as
the probability  ratio of finding  a WDMS binary containing  a He-core
white dwarf  compared to  finding a WDMS  binary containing  a CO-core
white dwarf within the SDSS follow-up spectroscopy.  As expected, this
bias is only mild for  early-type companions, where the secondary star
dominates the $i$-band light of the system~--~with the exception of M0
companions,   where   the  CO-core   white   dwarfs   have  a   larger
$\Teff(\mathrm{min})$ than  the He-core,  and hence the  integral over
the  cooling age  is smaller.   For later  type companions,  the ratio
gradually evolves towards  that of single white dwarfs,  except for M8
and M9 companions,  where the larger size of  the He-core white dwarfs
results in  lower $\Teff(\mathrm{max})$ compared to  the CO-core white
dwarfs, and correspondingly lower integrals of $Vd\tau$.

Given that the bulk of the WDMS binaries analysed here have companions
with spectral types in the range M3--M4 \citep{schreiberetal10-1}, the
bias that  they contain  He-core white dwarfs  is $\sim1.5$,  which is
\textit{lower}  than the  bias  for single  white dwarfs,  $\sim2.75$.
This implies slightly revised low-mass white dwarf fractions among our
WDMS binaries.  Considering equation\,\ref{eq-wide} and \,\ref{eq-all}
but  taking   into  account  the   additional  bias,  we   provide  in
Table\,\ref{t-final} the  best estimates for the  low-mass white dwarf
fractions among  our WDMS  binaries.  Note that  the most  likely PCEB
fractions for WDMS binaries containing low- and high-mass white dwarfs
($f_{\mathrm{PCEB, l}}$  and $f_{\mathrm{PCEB,  h}}$), as well  as the
fraction   of  WDMS   binaries  containing   high-mass   white  dwarfs
($f_{\mathrm{\Mwd>0.5}}$)    remain    the    same   as    given    in
Table\,\ref{t-stats}.

\begin{table}
\centering
\caption{\label{t-final}   Taking  into   account   selection  effects
  intrinsic  to SDSS  we provide  here  both the  revised fraction  of
  low-mass white  dwarfs among WDMS  binaries $f_{\mathrm{\Mwd<0.5}}$,
  as well as the revised  fraction of low-mass white dwarfs among wide
  WDMS  binaries  $f_{\mathrm{WIDE,c}}$.    The  ranges  provided  are
  obtained  by  assuming  different  values  of $sf$,  as  defined  in
  Table\,\ref{t-stats}  and  Sect.\,\ref{s-wideor}.   The most  likely
  PCEB fractions for WDMS binaries containing low- and high-mass white
  dwarfs ($f_{\mathrm{PCEB, l}}$  and $f_{\mathrm{PCEB, h}}$), as well
  as the  fraction of WDMS binaries containing  high-mass white dwarfs
  ($f_{\mathrm{\Mwd>0.5}}$)     remain      the     same     as     in
  Table\,\ref{t-stats}.   The revised  fractions of  all  low-mass and
  high-mass white dwarfs (those in  WDMS binaries in addition to those
  white dwarfs  that are apparently  single) formed in  close binaries
  (denoted here as $f_{\mathrm{CLOSE, l}}$ and $f_{\mathrm{CLOSE, h}}$
  respectively,  see  Sect.\,\ref{s-are})   are  also  provided.   For
  completeness we  also give the corrected fraction  of low-mass white
  dwarfs   among  single   white   dwarfs,  $f_{\mathrm{lmwd,   s}}$.}
\setlength{\tabcolsep}{1.5ex}
\begin{tabular}{ccccc}
\hline
\hline
 $f_{\mathrm{\Mwd<0.5}}$ &  $f_{\mathrm{WIDE,c}}$ & $f_{\mathrm{CLOSE,  l}}$ & $f_{\mathrm{CLOSE, h}}$ & $f_{\mathrm{lmwd, s}}$ \\
\hline
0.20-0.23 & 0.005-0.10 & 0.72-0.88 & 0.27-0.30 & 0.37 \\
\hline
\end{tabular}
\end{table}

A direct consequence  of the selection effects outlined  above is that
the  low-mass peak  in  the  white dwarf  mass  distribution of  PCEBs
(Fig.\,\ref{f-wdmassdist})   is  expected   to  be   less  pronounced.
However, the detected excess of low-mass white dwarfs in PCEBs remains
as both  close and wide WDMS  binaries are affected by  the same bias.
On the  other hand, comparing  white dwarf mass distributions  of wide
WDMS binaries and single white dwarfs is somewhat affected as the bias
towards the detection  of low-mass white dwarfs in  both subsamples is
slightly  different.  This may  partly explain  the relatively  low KS
probability of $4$ per cent  between the two distributions obtained in
Sect.\,\ref{s-comp}.

\subsection{A note on resolved binaries on SDSS images}

As discussed  in more  detail in \citet{schreiberetal10-1},  very wide
and  not too  distant  WDMS binaries  can  be resolved  on their  SDSS
images, and are  herefore excluded from our WDMS  binary sample.  This
implies that  we slightly overestimated the  PCEB fractions.  However,
this effect should  not dramatically change our results,  i.e.  by not
more than $5-10$ per cent \citep{schreiberetal10-1}.

\section{Discussion}
\label{s-discuss}

The main results  found in the previous sections  can be summarized as
follows.   White  dwarf mass  distributions  of  PCEBs  and wide  WDMS
binaries differ significantly. While the former contains a much larger
fraction of  low-mass white  dwarfs, the latter  is similar  to single
white  dwarf  mass  distributions.   This  is  the  so  far  strongest
observational confirmation that large numbers of low-mass white dwarfs
are formed due to binary  interactions.  However, even if we take into
account observational  biases and selection  effects, $\sim$0.5-10 per
cent  of the  wide WDMS  binary  sample also  contains low-mass  white
dwarfs  of   unclear  origin.   In   the  following  we   discuss  the
implications  of these results  on white  dwarf mass  distributions in
general and the  origin of apparently single low-mass  white dwarfs in
particular.

\subsection{Wide WDMS binaries as proxies for single white dwarfs}
\label{s-disc1}

We  have shown that  the mass  distribution of  wide WDMS  binaries is
similar to that  of single white dwarfs.  On the  one hand, this might
seem  trivial to  interpret as  the primary  in a  wide  binary should
evolve into  a white dwarf  the same way  a single star does.   On the
other hand,  the initial mass function  (IMF) of the  primary stars in
wide  binaries  and  the  IMF  of single  stars  are  not  necessarily
identical.   For  example,  the  binary  fraction  of  low-mass  (i.e.
$M\lappr1\,\Msun$) main sequence stars seems  to depend on the mass of
the primary star \citep{lada06-1}.   However, here we need to consider
only binary stars with primaries $\gappr1\,\Msun$, as lower mass stars
are still  on the main  sequence and have  not yet evolved  into white
dwarfs.   The  measured  binary  fractions  for solar  type  stars  is
$\sim60-70$  per cent \citep[e.g.][]{duquennoy+mayor91-1}  and similar
fractions     are      obtained     for     higher      mass     stars
\citep{dawson+schreoder10-1}.   Hence, it is  not surprising  that the
white  dwarf mass  distribution of  wide WDMS  binaries  resembles the
single white dwarf mass distribution and  we may use it as a proxy for
the  single  white  dwarf   mass  distribution.   In  particular,  the
$\sim0.5-10$  per  cent  of  low-mass  white dwarfs  among  wide  WDMS
binaries (Table\,\ref{t-final})  can be compared with  the fraction of
single low-mass white dwarfs.

\subsection{The origin of low-mass white dwarfs in wide binaries}
\label{s-disc2}

As mentioned  in the introduction,  a second peak  at $\sim0.4\,\Msun$
containing  about $10$  per cent  ($\sim4$ per  cent if  we  take into
account selection  effects in  SDSS, see Table\,\ref{t-final})  of the
systems  has  been  frequently   found  in  single  white  dwarf  mass
distributions \citep[e.g.][note  that this peak is not  obvious to the
  eye in Fig.\,\ref{f-wdmassdist2}, gray distribution, due to the size
  of    the    binning   used]{bergeronetal92-1,    bragagliaetal95-1,
  liebertetal05-1, kepleretal07-1}.  Extensive radial velocity surveys
have shown that about the half of these apparently single white dwarfs
are  in  fact   double  degenerates  \citep[see  e.g.][and  references
  therein]{kilicetal07-1}.   The remaining  $\sim2$ per  cent however,
represent a population  of low-mass white dwarfs that  do not show any
signs for a companion, i.e.  no infrared excess and no radial velocity
variations      \citep[see][]{maxtedetal00-3,     napiwotzkietal07-1}.
Possible explanations for the  existence of these stars include severe
mass-loss  on the  first giant  branch  \citep{kilicetal07-1}, stellar
envelope     ejection     due      to     nearby     giant     planets
\citep{nelemans+tauris98-1},  the merging of  two very  low-mass white
dwarfs   \citep{hanetal02-1,iben+tutokov86-2,iben90-1},  or  supernova
type Ia explosions in semi-detached  close binaries that blow away the
envelope   of  the   companion   thus  exposing   its  low-mass   core
\citep{justhametal09-1}.

Interestingly, the  fraction of low-mass white dwarfs  among wide WDMS
binaries  estimated in  Sect.\,\ref{s-seleff}, i.e.   $\sim0.5-10$ per
cent, resembles the typical $\sim4$  per cent of low-mass white dwarfs
among apparently single white  dwarfs.  Given that hierarchical triple
systems are  a common configuration  \citep[e.g.][]{tokovinin97-1}, it
appears plausible  to assume that  a significant fraction of  the wide
binaries containing a low-mass white dwarf are in fact triple systems,
or  descendants from  triple systems,  either containing  close double
degenerate primaries,  or low-mass white  dwarfs that formed  from the
merging  of two  very low-mass  white  dwarfs.  Indeed,  a handful  of
triple systems containing a  short-period double degenerate are known:
WD\,1704+481  has   been  identified  by   \citet{maxtedetal00-1}  and
\citet{finley+craig01-1} as a  triple-degenerate, containing a He-core
plus CO-core close binary double  degenerate with an orbital period of
0.1448\,d  and a  wide,  average-mass CO-core  white dwarf  companion.
WD\,1824+040  (G21--15)  is  a  similar  triple,  containing  a  close
double-degenerate   \citep[see][]{safferetal98-1}   with  a   6.266\,d
orbital period  \citep{maxted+marsh99-1} and a wide,  cool white dwarf
companion \citep{farihietal05-1}. Even more relevant to the discussion
here      are     PG\,1204+450,     a      close     double-degenerate
\citep{safferetal98-1} with a  period of 1.6\,d \citep{maxtedetal02-2}
and a  wide M4 companion \citep{farihietal05-1},  and PG\,1241--010, a
close double-degenerate with  a period of 3.3\,d \citep{marshetal95-1}
and a  wide M9 companion \citep{farihietal05-1}.   Therefore, a radial
velocity survey  of the  low-mass white dwarfs  among the  wide binary
candidates in our sample will be an interesting exercise.  In the same
way, as the double degenerate  scenario may account for low-mass white
dwarfs  in  wide WDMS  binaries,  the  formation  channel proposed  by
\citet{nelemans+tauris98-1} (stellar  envelope ejection due  to nearby
giant  planets) should work  in wide  binary systems.   Nearby planets
exist around the stellar  components of binary systems \citep[see][and
  references  therein]{desidera+barbieri07-1} and  the  presence of  a
wide companion does not affect  the interaction between the planet and
its  host star.   Another  possibility  is given  by  the single  star
formation   scenario   for   low-mass   white   dwarfs   proposed   by
\citet{kilicetal07-1}.   If  severe  mass  loss  of  the  white  dwarf
progenitors  on the first  giant branch  accounts for  single low-mass
white dwarfs, it should be present in wide binaries too as the stellar
components of wide binaries evolve just like single stars.

In    contrast,    the    supernova    Ia    channel    proposed    by
\citet{justhametal09-1}  appears to  be unlikely  to  produce low-mass
white dwarf primaries in  wide binary systems.  The explosion supposed
to strip  off the  envelope should accelerate  the core of  the former
secondary star far beyond the  velocity required to escape from a wide
M-dwarf companion.

\subsection{A note on spectroscopic white dwarf masses}

Since large scale  surveys such as the SDSS or  the SPY \citep[the ESO
  SN Ia Progenitor  survey, see][]{napiwotzkietal01-1} provide spectra
of  thousands  of white  dwarf  stars,  fitting  atmosphere models  to
observed  spectra  has  become  the  most frequently  used  method  to
determine    white    dwarf   masses    \citep[e.g.][]{finleyetal97-1,
  liebertetal05-1,  kepleretal07-1}.  The white  dwarf masses  used in
this  work have  been obtained  with such  an algorithm  combined with
spectral   decomposition   methods  \citep{rebassa-mansergasetal10-1}.
Atmosphere model fits to optical spectroscopy are generally thought to
provide reliable  estimates of white dwarf masses,  with the exception
of  the well  known problems  concerning cold  white  dwarfs mentioned
above.  For this reason, we  have excluded white dwarfs with effective
temperatures  below $12000$\,K,  and  by common  standards, our  white
dwarf masses should be robust and reliable.

However, recent  work suggest that  there may be a  systematic problem
with    mass    determination     from    model    atmosphere    fits.
\citet{silvestrietal01-1}   presented   gravitational  redshift   mass
estimates for  a sample  of 41 white  dwarfs in common  proper motions
pairs and  noticed that  their mean mass  was slightly higher  than in
previous  studies. More  recently  \citet{falconetal10-1} reported  an
average  white dwarf  mass of  $0.647$\,\Msun\ for  449  non-binary DA
white  dwarfs from  SPY  determined also  from gravitational  redshift
measurements.   This  value   is  again  significantly  exceeding  (by
$\sim0.03-0.05\,\Msun$)  the mean  white dwarf  masses that  have been
obtained from model atmosphere fits to large samples.  A similar trend
has  been  found  by  \citet{tremblay+bergeron09-1}  who  incorporated
improved physics of the Stark  broadening of the Balmer lines in their
white dwarf  atmosphere models  and obtained a  mean white  dwarf mass
higher by $0.034$\,\Msun.  It seems hence that the previous work based
on atmosphere models may  have systematically underestimated the white
dwarf masses.

Our  white  dwarf  masses  were  determined based  on  the  models  of
\citet{koesteretal05-1}, which did not incorporate the new Balmer line
profiles  of  \citet{tremblay+bergeron09-1}, and  hence  there is  the
possibility  that our masses  are systematically  lower than  the true
white dwarf  masses by $0.03-0.05$\,\Msun.  This would  imply a slight
decrease  in the  relative numbers  of low-mass  white dwarfs  in wide
binaries as  the close binary  fraction is increasing  towards smaller
masses    below    the    mass    limit   of    $0.5$\,\Msun\,    (see
Fig.\,\ref{f-detec}).   However, this effect  can clearly  not explain
the existence of all alleged  low-mass white dwarfs that appear not to
be within close binaries.

\subsection{The missing very low-mass white dwarf primaries}
\label{s-verylow}

Recently, a number of  extremely low-mass white dwarfs ($<0.2\,\Msun$)
have   been  found   in  double   degenerates  \citep{liebertetal04-1,
  kawkaetal06-1,             kilicetal07-2,             marshetal10-1,
  kulkarni+vankerkwijk10-1},    including    the    first    eclipsing
double-degenerate  binary \citep{steinfadtetal10-1}.  In  contrast, we
do not find any good  candidate for such an ultra-low-mass white dwarf
within  the  entire WDMS  binary  sample.  

For PCEBs this  difference might be related to  the fact that
the formation of very low-mass white dwarfs requires the progenitor of
the  white dwarf  to  fill its  Roche-lobe  early on  the first  giant
branch.  While  systems with relatively  massive companions apparently
can   survive  the  resulting   common  envelope   evolution,  systems
containing  M-dwarf companions may  instead merge  (as they  must have
smaller  initial  binary  separations  and consequently  less  orbital
energy is available to expel the envelope of the primary).

In wide  WDMS binaries,  the lack of  extremely low-mass  white dwarfs
seems  to  indicate  that  a  triple system  consisting  of  a  double
degenerate  primary (of which  at least  one of  the components  is an
extremely low-mass  white dwarf) with  an M-dwarf secondary in  a wide
orbit  (Sect.\,\ref{s-disc2}), might  not be  a  common configuration.
One may finally  speculate that the majority of  all double degenerate
primaries containing  extremely low-mass  white dwarfs have  merged to
form wide WDMS binaries in which  the resulting white dwarf mass is at
least $\ga0.35\Msun$.

\section{Conclusion}
\label{s-concl}

The  white dwarf  mass  distributions  of PCEB  and  wide WDMS  binary
candidates are significantly different. While the mass distribution of
wide WDMS  binary candidates resembles  those of single  white dwarfs,
the  PCEB white  dwarf mass  distribution contains  a large  number of
low-mass white  dwarfs.  Taking into  account both the  PCEB detection
probabilities  of the measurements  and selection  effects in  SDSS we
find that the large majority of low-mass white dwarfs resides in close
binary stars.   This result confirms  a crucial prediction  of current
theories of close  binary evolution and provides the  so far strongest
observational  evidence  for common  envelope  evolution forming  most
close compact binary stars.

In agreement with  the fraction of $\sim4$ per  cent apparently single
low-mass  white  dwarfs  identified  in single  white  dwarf  samples,
$\sim0.5-10$  per cent  of the  wide binaries  in our  sample  seem to
contain low-mass  white dwarfs.  These  low-mass white dwarfs  in wide
binaries must have either formed due to exceptionally strong mass-loss
of  the  primary progenitor  on  the first  giant  branch  or are  the
descendants  of triple  systems, and  some are  expected to  contain a
close double degenerate primary component.

\section*{Acknowledgments.}

ARM acknowledges financial support from  Fondecyt in the form of grant
number 3110049,  ESO and Gemini/Conicyt  (32080023).  MRS acknowledges
support from Fondecyt (grants 1061199  and 1100782), and the Center of
Astrophysics  in  Valparaiso  (CAV).   We thank  Kepler  Oliveira  for
providing us his white dwarf  mass measurements and Detlev Koester for
helpful discussions.  We also  thank the anonymous referee for his/her
suggestions that helped improving the quality of the paper.


\label{lastpage}

\end{document}